\documentclass{IEEEtran}
\usepackage{graphicx}
\usepackage[fleqn]{amsmath}
\usepackage[varg]{txfonts}

\usepackage{latexsym}
\usepackage{amssymb}
\usepackage{bm}
\usepackage{subfigure}
\usepackage{algorithm}
\usepackage{algorithmic}
\usepackage{epic,eepic}

\newtheorem{theorem}{Theorem}
\newtheorem{lemma}{Lemma}
\newtheorem{corollary}{Corollary}
\newtheorem{proposition}{Proposition}
\newtheorem{remark}{Remark}

\newtheorem{example}{Example}

\renewcommand{\algorithmicrequire}{\textbf{Input:}}
\renewcommand{\algorithmicensure}{\textbf{Output:}}
\renewcommand{\mathtt}[1]{\texttt{#1}}

\newcommand{\bs}{\boldsymbol}
\newcommand{\neib}[1]{\mathcal{N}_{\mathrm{#1}}}
\newcommand{\ot}{\leftarrow}

\begin{document}

\title{
Zigzag Decodable Fountain Codes
}

\author{
Takayuki~Nozaki%
\thanks{
The material in this paper was presented in part at 
International Symposium on Information Theory and its Applications (ISITA2014) \cite{nozaki_zdf},
Melbourne, Australia, October 2014.

The author is with Division of Natural Science,
Yamaguchi University,
Yamaguchi 753-8511, Japan
e-mail: tnozaki@yamaguchi-u.ac.jp
}
}

\maketitle

\begin{abstract}
This paper proposes a fountain coding system which has lower space decoding complexity and lower decoding erasure rate than the Raptor coding systems.
The main idea of the proposed fountain code is employing shift and exclusive OR to generate the output packets.
This technique is known as the zigzag decodable code, which is efficiently decoded by the zigzag decoder.
In other words, we propose a fountain code based on the zigzag decodable code in this paper.
Moreover, we analyze the overhead, decoding erasure rate, decoding complexity, and asymptotic overhead of the proposed fountain code.
As the result, we show that the proposed fountain code outperforms the Raptor codes in terms of the overhead and decoding erasure rate.
Simulation results show that the proposed fountain coding system outperforms Raptor coding system in terms of the overhead and the space decoding complexity.
\end{abstract}

\section{Introduction}
On the Internet, each message is transmitted in a sequence of packets.
We consider that the packets which are not correctly received are erased.
Hence, the Internet is modeled as the {\it packet erasure channel} (PEC).
The erased packets are re-transmitted by the automatic repeat-request (ARQ) in the case of the transmission control protocol (TCP).
On the other hand, the sender cannot resend the packets in the case of user diagram protocol (UDP), which can be used in multicasting, broadcasting, and unicasting under the large round trip time.


{\it Fountain code} \cite{Fountain} realizes reliable communication for the UDP.
We assume that the transmitted message are divided into $k$ source packets.
Fountain code produces infinite output packets from $k$ source packets.
The receivers decode the message from {\it arbitrary} $k(1+\alpha)$ output packets with $\alpha \ge 0$.
Hence, the receivers need not request retransmitting packets.
The parameter $\alpha$ is referred to as {\it overhead for received packets}.

Luby first realized the concepts of the fountain code with LT codes
\cite{LT}.
Each output packet of the LT code is generated as follows.
Firstly, the encoder randomly chooses the degree $d$ of the output packet according to the degree distribution $\Omega(x)$.
Secondly, the encoder randomly chooses $d$ distinct source packets.
Finally, the encoder outputs bit-wise exclusive OR (XOR) of the $d$ source packets as an output packet.
Decoding of LT codes is similar to that of low-density parity-check (LDPC) codes over the binary erasure channel.
More precisely, the decoder constructs the factor graph from the received packets and recovers the source packets by using an iterative decoding algorithm, called {\it peeling algorithm} \cite{Luby97}.

Raptor codes \cite{Raptor} are fountain codes which achieves arbitrarily small $\alpha$ as $k \to \infty$ with linear time encoding and decoding.
Encoding of Raptor code is divided into two stages.
In the first stage, the encoder generates the {\it precoded packets} from the source packets by using an erasure correcting code.
In the second stage, the encoder generates the output packets from the precoded packets by using an LT code.
Decoding of the Raptor codes is similar to that of the LT codes.

The RaptorQ code \cite{luby2011raptorq,shokrollahi2011raptor} achieves small overhead $\alpha$ under maximum likelihood decoding, called {\it inactivation decoding}.
However, it is not suitable for the receivers with low performance processors since the inactivation decoder has higher complexity than most iterative decoders.
Moreover, since the RaptorQ code employs a dense non-binary linear code for the precode, the decoding complexity is high.
Hence, we focus on the fountain coding system under an iterative decoding algorithm.


Gollakota and Katabi \cite{ZZD_HT} proposed zigzag decoding to combat hidden terminals in wireless networks.
Sung and Gong \cite{ZZD_DS} proposed zigzag decodable (ZD) codes which are efficiently decoded by the zigzag decoder, for the distributed storage systems.
In recent years, ZD codes are applied to network coding \cite{6979934}. 
As a similar study, Qureshi {\it et al}.\ \cite{6275780} proposed triangular codes and back-substitution decoding method for the index decoding problem.
Both ZD codes and triangular codes generate output packets from the source packets by using shift and XOR.
Hence, the length of the output packets is slightly longer than that of the source packets.
However, when the length of source packets is large, the growth of overhead for the received bits is very small.
Qureshi {\it et al}.\ \cite{DBLP:journals/corr/abs-1305-0918} suggested that the triangular codes can be applied to the fountain codes.
However, there are no comparison with other fountain codes and there are no analysis of the fountain codes based on triangular coding.

In this paper, we investigate the fountain codes based on ZD coding.
The proposed fountain code is regarded to as a generalization of Raptor code.
More precisely, the proposed code generates output packets from precoded packets by using shift and XOR.
In this paper, as a first step of research, we compare the proposed fountain code with original Raptor code given in \cite{Raptor}.

The contributions of this paper are the followings: 
(1) We give factor graph representations of the ZD codes.
(2) We propose a fountain code based on ZD coding and its decoding algorithm.
(3) We prove that the decoding erasure probability of the proposed fountain coding system is lower than that for the Raptor coding system.
(4) We analyze the overhead for a large number of source packets by the density evolution.
Moreover, simulation results shows that the proposed fountain codes outperforms Raptor codes in terms of overhead and space complexity of decoding.
Nowadays, the proposed fountain coding system is extended to the limited memory case by the Jun et al.~\cite{7354590}.

The rest of the paper is organized as follows.
Section \ref{sec:tc} briefly explains the ZD codes and zigzag decoding by a toy example.
Section \ref{sec:fg-rep} gives factor graph representations of the ZD codes.
Section \ref{sec:fc-tc} proposes the fountain codes based on ZD coding and its decoding algorithm.
Section \ref{sec:pe} analyzes the overhead, decoding performance, and decoding complexity of the proposed fountain coding system.
Moreover, simulation results in Section \ref{sec:pe} give that the proposed fountain coding system outperforms Raptor coding system in terms of the overhead for the received packets.
Section \ref{sec:asympto} evaluates the overhead for the large number of source packets by deriving the density evolution equations for the proposed fountain coding system.
Section \ref{sec:conc} concludes this paper.

\section{Example of ZD Codes And Zigzag Decoding \label{sec:tc}}
This section explains the ZD code and the zigzag decoding algorithm with a toy example.
Moreover, we point out a drawback of the original zigzag decoding algorithm.

As a toy example, we consider a ZD code which generates two encoded packets from two source packets with length $\ell$.
The first encoded packet ${\bs x}_1 = (x_{1,1}, x_{1,2},\dots, x_{1,\ell})$ is generated from the bit-wise XOR of two source packets ${\bs s}_1 = (s_{1,1},s_{1,2},\dots,s_{1,\ell}), {\bs s}_2 = (s_{2,1},s_{2,2},\dots,s_{2,\ell})$.
The second encoded packet ${\bs x}_2 = (x_{2,1}, x_{2,2},\dots, x_{2,\ell+1})$ is generated from the bit-wise XOR of ${\bs s}_2$ with a right shift and ${\bs s}_1$.
After shifting the packet, zeros are filled.
Notice that the length of the second packet is $\ell +1$.
Figure \ref{fig:ex1} illustrates this ZD code.
Since the XOR is equivalent to the addition over the Galois field of order 2, we denote the XOR operation, by $+$.S

ZD codes are efficiently decoded by the zigzag decoding algorithm \cite{ZZD_HT,ZZD_DS}.
The zigzag decoding algorithm starts from the {\it left} of the packets.
In a similar way to the peeling decoding algorithm for the LDPC code over the binary erasure channel (BEC), the zigzag decoding algorithm proceeds by solving linear equations with one unknown variable.

For the ZD code in Fig.~\ref{fig:ex1}, the zigzag decoding algorithm proceeds as the following way.
The decoder recovers $s_{1,1}$ from $x_{2,1}$ since $s_{1,1} = x_{2,1}$.
The decoder recovers $s_{2,1}$ by solving $x_{1,1} = s_{1,1}+s_{2,1} = x_{2,1}+s_{2,1}$.
Similarly, the decoder recovers $s_{1,2}, s_{2,2},\dots, s_{2,\ell}$ and decoding is success.

\begin{remark} \label{rem:1} \upshape
Recall that the original zigzag decoding algorithm \cite{ZZD_HT,ZZD_DS} starts from the {\it left} of the encoding packets.
Hence, the ZD code described as in Fig.~\ref{fig:ex2} is not decoded by the original zigzag decoding algorithm.
However, if decoding starts from the {\it right} of the encoding packets, the ZD code in Fig.~\ref{fig:ex2} is decodable.
Actually, $s_{i,\ell}$ is recoverable from $x_{i,\ell+1}$ for $i=1,2,3$.
Substituting these values, we get $s_{i,\ell-1}$ for $i=1,2,3$ in a similar way.
Finally, we get $s_{i,1}$ for $i=1,2,3$ and decoding is success.

Hence, the zigzag decoding algorithm is improved if decoding starts from the {\it left and right} of the encoding packets.
This idea had been proposed in the literature of the triangular code \cite{6275780}.
We also use this technique in Section \ref{sec:fc-tc}.
\end{remark}

\begin{remark} \upshape
Similar to the ZD codes, the triangular codes \cite{6275780} generate the encoded packets from the source packets by using shift and XOR.
The triangular code chooses distinct shift amounts of the source packets.
Hence, the triangular code is always decodable from the left of the encoded packets.
This decoding algorithm is referred to as back-substitution algorithm \cite{6275780}. 
Since there are no constraints on the shift amounts of source packets for the ZD codes, the triangular codes are a special case of the ZD codes.
\end{remark}

\begin{figure}[tb]
\begin{center}
\begin{picture}(241,40)
\begin{scriptsize}
 \put(10,  25){\framebox(20,9){$s_{1,1}$}}
 \put(30.5,25){\framebox(20,9){$s_{1,2}$}}
 \put(51,  25){\framebox(33.5,9){$\cdots$}}
 \put(85,  25){\framebox(20,9){$s_{1,\ell}$}}

 \put(0, 18){$+$}

 \put(10,  15.5){\framebox(20,9){$s_{2,1}$}}
 \put(30.5,15.5){\framebox(20,9){$s_{2,2}$}}
 \put(51,  15.5){\framebox(33.5,9){$\cdots$}}
 \put(85,  15.5){\framebox(20,9){$s_{2,\ell}$}}

 \put(10,  0){\framebox(20,9){$x_{1,1}$}}
 \put(30.5,0){\framebox(20,9){$x_{1,2}$}}
 \put(51,  0){\framebox(33.5,9){$\cdots$}}
 \put(85,  0){\framebox(20,9){$x_{1,\ell}$}}
 \put(125,  25){\framebox(20,9){$s_{1,1}$}}
 \put(145.5,25){\framebox(20,9){$s_{1,2}$}}
 \put(166,  25){\framebox(33.5,9){$\cdots$}}
 \put(200,25){\framebox(20,9){$s_{1,\ell}$}}
 \put(220,25){\makebox(20,9){$0$}}

 \put(115, 18){$+$}

 \put(125,  15.5){\makebox(20,9){$0$}}
 \put(145.5,15.5){\framebox(20,9){$s_{2,1}$}}
 \put(166,15.5){\framebox(33.5,9){$\cdots$}} 
 \put(200,15.5){\framebox(20,9){\hspace{1pt}$s_{2,\ell\hspace{-0.5pt}-\hspace{-0.5pt}1}$}}
 \put(220.5,15.5){\framebox(20,9){$s_{2,\ell}$}}

 \put(125,  0){\framebox(20,9){$x_{2,1}$}}
 \put(145.5,0){\framebox(20,9){$x_{2,2}$}}
 \put(166,  0){\framebox(33.5,9){$\cdots$}}
 \put(200,0){\framebox(20,9){$x_{2,\ell}$}}
 \put(220.5,0){\framebox(20,9){\hspace{1pt}$x_{2,\ell\hspace{-0.5pt}+\hspace{-0.5pt}1}$}}
\end{scriptsize}
\end{picture}
\end{center}
\caption{A toy example of ZD code \label{fig:ex1}}
\end{figure}
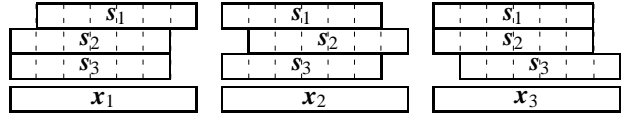
\begin{figure}[tb]
 \begin{center}
  \begin{picture}(230,40)
   \put(10,31.5){\framebox(60,9){${\bs s}_1$}}
   \put(0 ,22  ){\framebox(60,9){${\bs s}_2$}}
   \put(0 ,12.5){\framebox(60,9){${\bs s}_3$}}
   \put(0 ,0   ){\framebox(70,9){${\bs x}_1$}}
   \dashline{2}(9.5,12.5)(9.5,40.5)
   \dashline{2}(20,12.5)(20,40.5)
   \dashline{2}(30,12.5)(30,40.5)
   \dashline{2}(40,12.5)(40,40.5)
   \dashline{2}(50,12.5)(50,40.5)
   \dashline{2}(60.5,12.5)(60.5,40.5)

   \put(80,31.5){\framebox(60,9){${\bs s}_1$}}
   \put(90,22  ){\framebox(60,9){${\bs s}_2$}}
   \put(80,12.5){\framebox(60,9){${\bs s}_3$}}
   \put(80,0   ){\framebox(70,9){${\bs x}_2$}}
   \dashline{2}(90,12.5)(90,40.5)
   \dashline{2}(100,12.5)(100,40.5)
   \dashline{2}(110,12.5)(110,40.5)
   \dashline{2}(120,12.5)(120,40.5)
   \dashline{2}(130,12.5)(130,40.5)
   \dashline{2}(140,12.5)(140,40.5)

   \put(160,31.5){\framebox(60,9){${\bs s}_1$}}
   \put(160,22  ){\framebox(60,9){${\bs s}_2$}}
   \put(170,12.5){\framebox(60,9){${\bs s}_3$}}
   \put(160,0   ){\framebox(70,9){${\bs x}_3$}}

   \dashline{2}(170,12.5)(170,40.5)
   \dashline{2}(180,12.5)(180,40.5)
   \dashline{2}(190,12.5)(190,40.5)
   \dashline{2}(200,12.5)(200,40.5)
   \dashline{2}(210,12.5)(210,40.5)
   \dashline{2}(220.5,12.5)(220.5,40.5)

  \end{picture}
 \end{center}
 \caption{A ZD code which is not decoded by the original zigzag decoding algorithm \label{fig:ex2}}
\end{figure}

\section{Factor Graphs for ZD codes \label{sec:fg-rep}}
This section explains a matrix representation of a ZD code \cite{ZZD_DS} and presents factor graph representations of a ZD codes.

\subsection{Matrix Representation for ZD codes \cite{ZZD_DS}}
Let $\ell$ be the length of source packets.
Denote the number of source packets, by $k$.
A polynomial representation of the $i$-th source packet $(s_{i,1},s_{i,2},\dots, s_{i,\ell})$ is defined as
\begin{equation*}
 s_i(z) 
  =
 {\textstyle \sum_{j=1}^{\ell}} s_{i,j} z^j.
\end{equation*}
Then, for the ZD codes, the polynomial representation of the $i$-th encoded packet is given by
\begin{equation}
 x_i(z) = 
  {\textstyle \sum_{j=1}^k} g_{i,j}(z) s_j(z), \label{eq:enc1}
\end{equation}
where $g_{i,j}(z)$ is a monomial of $z$, i.e, $g_{i,j}(z) \in \{0,1,z,z^2,\dots\}$.
We denote the degree of $g_{i,j}(z)$, by $\deg (g_{i,j})$.
Then, the length of the $i$-th encoded packet is $\ell + \max_{j}\deg (g_{i,j})$.
We get the following matrix representation with Eq.~\eqref{eq:enc1}:
\begin{equation*}
 {\bs x}(z) = {\bs G}(z) {\bs s}(z).
\end{equation*}

\begin{example} \upshape
The matrix representation of the ZD code in Fig.~\ref{fig:ex1} is 
\begin{equation*}
{\bs G}(z) = \begin{pmatrix}1 & 1 \\ 1 & z\end{pmatrix}.
\end{equation*}
\end{example}

\subsection{Factor Graph Representation of ZD codes \label{ssec:fg_zd}}
In this section, we give packet-wise and bit-wise factor graph representations of the ZD codes.

Firstly, we give a packet-wise factor graph representation of a ZD codes.
The factor graphs of the ZD codes consist of the sets of nodes $\mathsf{V}_{\mathrm{s}}, \mathsf{V}_{\mathrm{x}}, \mathsf{C}$ and {\it labeled} edges.
The nodes in $\mathsf{V}_{\mathrm{s}}, \mathsf{V}_{\mathrm{x}}$ represent source packets and encoded packets, and are called source nodes and encoded nodes, respectively.
The number of source packets (resp.\ encoding packets) is equal to $|\mathsf{V}_{\mathrm{s}}|$ (resp.\ $|\mathsf{V}_{\mathrm{x}}|$).
The nodes in $\mathsf{C}$ represent constraints for the neighbor nodes, and are called factor nodes.
The number of nodes in $\mathsf{C}$ is equal to $|\mathsf{V}_{\mathrm{x}}|$.
All the encoded nodes are of degree one and the $i$-th encoded node and the $i$-th factor node are connected by an edge labeled by $1$.
If $g_{i,j}(z) \neq 0$, then the $j$-th source node and the $i$-th factor node are connected by an edge labeled by $g_{i,j}(z)$.
If $g_{i,j}(z) = 0$, then the $j$-th source node and the $i$-th factor node are not connected.
Note that the $i$-th factor node represents a constraint such that 
\begin{align}
 & {\textstyle \sum_{j\in\neib{c}(i)}} g_{i,j}(z)s_j(z) = x_i(z), \notag \\
\iff&
 {\textstyle \sum_{j\in\neib{c}(i)}} 
      s_{j, t - \deg(g_{i,j})} = x_{i,t} ~~\forall t , \label{eq:constraint}
\end{align}
where $\neib{c}(i)$ gives the set of indexes of the source nodes connecting to the $i$-th factor node.

Secondly, we give a bit-wise factor graph representation of the ZD codes.
In this representation, edges are not labeled.
Each source node and each encoded node corresponds to a bit of source packets and encoded packets, respectively.
Each factor node in this representation gives a constraint such that XOR of the bits corresponding to the neighbor source nodes and the neighbor encoded node is 0.
The $(j,t)$-th source node and $(i,t+t')$-th factor node are connected by an edge if $g_{i,j} = z^{t'}$.
The $(i,t)$-th factor node and $(i,t)$-th encoded node are also connected by an edge.

\begin{example} \upshape
Figures \ref{fig:fg1_pw} and \ref{fig:fg1_bw} show the packet-wise and bit-wise factor graph representations of the ZD code in Fig.~\ref{fig:ex1} with $\ell = 4$, respectively.
The black circle, white circle, and white square represent source nodes, encoded nodes and factor nodes, respectively.
Each dashed rectangular in Fig.~\ref{fig:fg1_bw} corresponds to a node in Fig.~\ref{fig:fg1_pw}.



\begin{figure}[tb]
  \begin{center}
  \subfigure[Packet-wise representation]{%
   \begin{picture}(100,80)(-15,0)
    \put(15,0){\includegraphics[height=80pt]{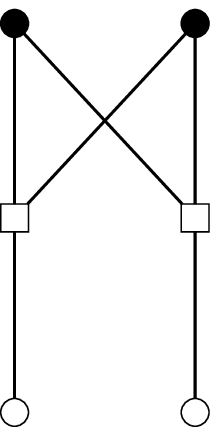}}
    \put(  0,  2){$\mathsf{v}_{\mathsf{x},1}$}
    \put( 60,  2){$\mathsf{v}_{\mathsf{x},2}$}
    \put(  0, 38){$\mathsf{c}_{1}$}
    \put( 60, 38){$\mathsf{c}_{2}$}
    \put(  0, 73){$\mathsf{v}_{\mathrm{s},1}$}
    \put( 60, 73){$\mathsf{v}_{\mathrm{s},2}$}

    \put( 12, 20){$1$}
    \put( 55, 20){$1$}

    \put( 12, 60){$1$}
    \put( 23, 70){$1$}
    \put( 40, 70){$1$}
    \put( 55, 60){$z$}
   \end{picture}
    \label{fig:fg1_pw}}
  \hspace{5mm}
  \subfigure[Bit-wise representation]{%
    \begin{picture}(100,80)
      \put(10,0){\includegraphics[height=80pt]{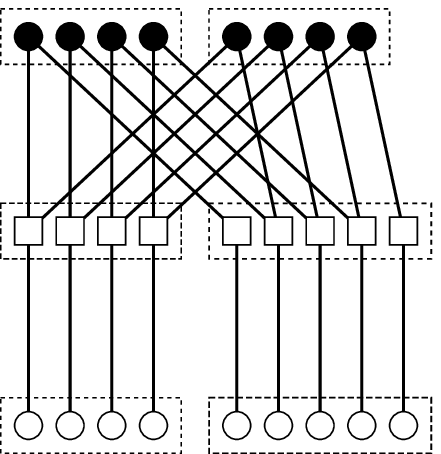}}
    \end{picture}
    \label{fig:fg1_bw}}
  \caption{ Factor graph representation of the ZD code in Fig.~\ref{fig:ex1} \label{fig:fg1}}
 \end{center}
\end{figure}

\end{example}

\begin{remark} \label{rem:3}
Decoding of ZD codes is regarded to as recovering the source packets from the encoded packets.
In the bit-wise factor graph representation, this problem is transformed into recovering black circles from the white circles under the constraints given by the squares.
In the terminology of LDPC codes, the black (resp.\ white) circles represent punctured (resp.\ transmitted) bits.
In this case, since transmitted bits have no errors, the punctured bits are recovered by peeling algorithm  \cite{Luby97}.
The peeling algorithm over the bit-wise factor graph is regarded as the improved zigzag decoding, given in Remark \ref{rem:1}.
\end{remark}

\section{Fountain Code Based on ZD Coding \label{sec:fc-tc}}
In this section, we propose a fountain coding system based on ZD coding.
In a similar way to Raptor codes, the proposed fountain code firstly generates precoded packets from source packets by using an LDPC code.
Next the proposed fountain code generates the output packets from the precoded packets with inner coding, which is a combination of an LT code and a ZD code.
Moreover, this section gives a decoding algorithm for the proposed fountain codes.

\subsection{Encoding}
The system parameters for the proposed fountain coding system are the precode $\mathcal{C}$, the degree distribution of the inner code $\Omega(x)=\sum_{i}\Omega_i x^i$ and the {\it shift distribution} $\Delta(x)=\sum_{i=0}^{s_m}\Delta_i x^i$, where $\Delta_i$ represents the probability that the shift amount is $i$.
Notice that $\Omega(1) = 1$ and $\Delta(1) = 1$.

Similarly to the Raptor codes, the proposed fountain code generates the precoded packets $({\bs a}_1,\dots, {\bs a}_n)$ from the source packets $({\bs s}_1,\dots, {\bs s}_{k})$ by the precode $\mathcal{C}$ in the first stage.
In the second stage, the proposed fountain code generates the infinite output packets as the following procedure for $t=1,2,\dots$.
\begin{enumerate}
\item
Choose a degree $d$ of the $t$-th output packet according to the degree distribution $\Omega(x)$. In other words, choose $d$ with probability $\Omega_d$.
\item \label{stp:enc2}
Choose $d$-tuple of shift amounts $(\tilde{\delta}_1,\dots,\tilde{\delta}_d)\in[0,s_m]^d$ in independent of each other according to shift distribution $\Delta(x)$, where, $[a,b]$ denote the set of integers between $a$ and $b$. 
Define $\tilde{\delta}_{\mathrm{min}} := \min_{i\in[1,d]} \tilde{\delta}_i$ and calculate $\delta_i := \tilde{\delta}_i - \tilde{\delta}_{\mathrm{min}}$ for $i\in[1,d]$.
\item
Choose $d$ distinct precoded packets uniformly.
Let $(j_1,j_2,\dots,j_d)$ denote the $d$-tuple of indexes of the chosen precoded packets.
Then the polynomial representation for the $t$-th output packet is given as
\begin{equation*}
 {\textstyle \sum_{i=1}^{d}} z^{\delta_i} a_{j_i}(z).
\end{equation*}
Note that the information of the tuples $(\delta_1,\dots,\delta_d)$, $(j_1,\dots,j_d)$ is in the header of the $t$-th output packet.
\end{enumerate}

In Step \ref{stp:enc2}, the encoding algorithm normalizes the shift amount, i.e, set $\delta_i = \tilde{\delta}_i - \tilde{\delta}_{\mathrm{min}}$.
Unless the algorithm normalize the shift amount, the output packet contains extra $\tilde{\delta}_{\mathrm{min}}$ bits in the head of the packet.
Hence, this normalization avoids extra bits in the output packets.

\begin{remark}
The length of the output packets is slightly longer than that of the source packets.
Denote the length of the $t$-th output packet, as $\ell+\ell_t$, with $\ell_t \ge 0$.
For the $t$-th output packet, assume that the degree is chosen as $d$ and the $d$-tuple of shift amounts is chosen as $(\tilde{\delta}_1,\dots, \tilde{\delta}_d)$.
Since the length of the $t$-th output packet is described as $\ell + \max_{i\in[1,d]} \deg (z^{\delta_i})$,
the additional length of the $t$-th output packet $\ell_t$ is
\begin{equation}
 \ell_t 
 = \max_{i\in[1,d]} \delta_i
  =
  \max_{i\in[1,d]} \tilde{\delta}_i - \min_{i\in[1,d]} \tilde{\delta}_i \label{eq:leng}
\end{equation}
\end{remark}

\subsection{Decoding}
Let $ {\bs r}_1,{\bs r}_2,\dots, {\bs r}_{\tilde{k}} $ be $\tilde{k}$ received packets for a receiver, where $\tilde{k} = k(1+\alpha)$.
Firstly, similar to the Raptor code, the decoder of the proposed fountain coding system constructs a factor graph from the precode $\mathcal{C}$ and headers of the received packets.
The generated factor graphs depend on receivers, since the $\tilde{k}$ received packets depend on receivers.
After constructing a factor graph, the decoder recovers source packets from $\tilde{k}$ received packets in a similar way to the peeling decoder for the LDPC code over the BEC.
In the following, we explain the construction of a factor graph for a ZD fountain code.

\subsubsection{Construction of Factor Graph}
In this section, we give the factor graph for the ZD fountain code.
Roughly speaking, the factor graph for the ZD fountain code is represented 
by a concatenation of the Tanner graph of the precode and the factor graph of the inner code.

Firstly, we explain the packet-wise representation of a factor graph.
The factor graphs for the ZD fountain code contains of labeled edges and the four kinds of nodes:
$n$ variable nodes representing precoding packets $\mathtt{V}_{\rm p} = \{\mathtt{v}_1, \dots, \mathtt{v}_{n}\}$,
$m$ check nodes on the precode code $\mathtt{C} = \{\mathtt{c}_1, \dots, \mathtt{c}_{m}\}$,
$\tilde{k}$ variable nodes representing received packets $\mathtt{V}_{\rm r} = \{\mathtt{v}_1', \dots, \mathtt{v}_{\tilde{k}}'\}$,
and $\tilde{k}$ factor nodes on the inner code $\mathtt{F} = \{ \mathtt{f}_1, \dots, \mathtt{f}_{\tilde{k}} \}$.
Note that $m:= n-k$.
The edge connection between $\mathtt{V}_{\rm p}$ and $\mathtt{C}$ is decided from the parity check matrix ${\bf H}$ of the precode.
More precisely, $\mathtt{c}_i$ and $\mathtt{v}_j$ are connected to an edge labeled by $1$ if and only if the $(i,j)$-th entry of ${\bf H}$ is equal to $1$.
The edge connection between $\mathtt{V}_{\rm p}$ and $\mathtt{f}_i$ is decided from the header of the $i$-th received packet.
If the header of the $i$-th received packet represents $(\delta_1,\dots, \delta_{d})$ and $(j_1, \dots, j_d)$,
an edge labeled by $z^{\delta_k}$ connects $\mathtt{f}_i$ and $\mathtt{v}_{j_k}$ for $k\in[1,d]$.
We denote the label on the edges connecting to $\mathtt{f}_i$ and $\mathtt{v}_{j}$, by $z^{\delta_{i,j}}$.
For $i\in[1,\tilde{k}]$, an edge connects $\mathtt{f}_i$  and $\mathtt{v}'_i$.

Denote the set of indexes of the variable nodes in $\mathtt{V}_{\rm p}$ adjacent to the $i$-th check node $\mathtt{c}_i$ (resp.\ $j$-th factor node $\mathtt{f}_j$),
by $\mathcal{N}_{\mathtt{c}}(i)$ (resp.\ $\mathcal{N}_{\mathtt{f}}(j)$).
Similarly, denote the set of indexes of the check nodes (resp.\ factor nodes) adjacent to the $i$-th variable node $\mathtt{v}_i$,
by $\mathcal{N}_{\mathtt{vc}}(i)$ (resp.\ $\mathcal{N}_{\mathtt{vf}}(i)$).
Then, the $i$-th check node $\mathtt{c}_i$ and the $j$-th factor node $\mathtt{f}_j$ gives the following constraints
\begin{align*}
&{\textstyle \sum_{k\in \mathcal{N}_{\mathtt{c}}(i)} a_{k}(z) = 0, 
\quad\text{and}\quad \sum_{k\in \mathcal{N}_{\mathtt{f}}(j)} z^{\delta_{k,j}} a_k(z) = r_j(z)},
\end{align*}
respectively.

The bit-wise representation of a factor graph is obtained by a similar way in Section \ref{ssec:fg_zd}.

\begin{example}
Figure \ref{fig:foun} illustrates an example of a factor graph in the packet-wise representation.
In this example, we employ a $(2,4)$-regular LDPC code as the precode.
The black (resp.\ white) circles are the variable nodes representing the precoded (resp.\ received) packets.
The black and white squares represent the check nodes of the precode and factor nodes of the inner code, respectively.
Each edge is labeled by a monomial of $z$.
Note that all the edges in the factor graph corresponding to the precode are labeled by $1$.

\begin{figure}[tb]
\begin{center}
 \begin{picture}(140,90)
 \begin{small}
 \put(0,0){\includegraphics[height = 90pt]{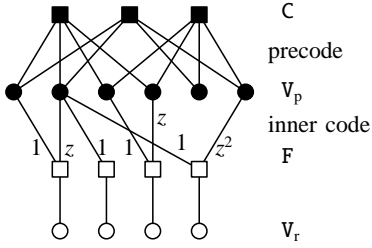}}

 \put(10,33){1}
 \put(23,33){$z$}

 \put(35,33){1}

 \put(45,33){1}
 \put(58,45){$z$}

 \put(65,35){$1$}
 \put(80,33){$z^2$}

 \put(100, 70){{\small precode}}
 \put(100, 42){{\small inner code}}
 \put(105, 85){$\mathtt{C}$}
 \put(105, 55){$\mathtt{V}_{\mathrm p}$}
 \put(105, 30){$\mathtt{F}$} 
 \put(105,  2){$\mathtt{V}_{\mathrm r}$}
 \end{small}
 \end{picture}
\end{center}
\caption{An example of a factor graph for a proposed fountain code. \label{fig:foun}}
\end{figure}

\end{example}

\subsubsection{Peeling Algorithm}
In this section, we explain decoding algorithm based on peeling algorithm (PA) for the ZD fountain codes.
As shown in Remark \ref{rem:3}, we can decode the proposed fountain codes by using the PA over the bit-wise factor graph representation.
However, to understand the difference between Raptor code and proposed fountain code, we present a two stage decoding algorithm, which is the concatenation of packet-wise PA and bit-wise PA.

In the PA, all the check nodes and factor nodes have memories.
At each iteration, PA updates the memories in the nodes and the {\it residual graph}, which consists of the un-recovered variable nodes, edges connecting to those variable nodes, and the check nodes which is adjacent to those variable nodes.

In the packet-wise PA, the memory length of each check node depends on the packet length.
More precisely, the memory length of the all check nodes in $\mathtt{C}$ is $\ell$ and
the memory length of factor node $\mathtt{f}_{t} \in \mathtt{F}$ is equal to the length of $t$-th received packet ${\bs r}_t$.
We denote the polynomial representation of the memory value in $\mathtt{c}_i$ (resp.\ $\mathtt{f}_i$), by $w_i(z)$ (resp.\ $w_i'(z)$).

The details of packet-wise PA is described in Algorithm \ref{algc:1}.
In Algorithm \ref{algc:1}, $\tau$ represents the decoding round and $*$ stands erased symbol.
Algorithm \ref{algc:1} stops if there is not any factor nodes and check nodes of degree 1 in the residual graph $\mathtt{G}$.
If Algorithm \ref{algc:1} outputs ${\bs a}_j \neq *^{\ell}$ for all $j\in[1,n]$, the decoding succeeds.
Otherwise, execute the bit-wise PA since there are some un-recovered precoded packets.

Before the bit-wise PA, decoder transforms the residual graph $\mathtt{G}$ into the bit-wise representation $\mathtt{G}_{\mathrm{b}}$ in a similar way to the generation of bit-wise factor graph.
In the bit-wise representation, $\mathtt{v}_j\in \mathtt{V}_{\mathrm{p}}$ (resp.\ $\mathtt{c}_i\in\mathtt{C}_{\mathrm{p}}$) in $\mathtt{G}$ is replaced with $\ell$ nodes $\mathtt{v}_{j,1},\dots,\mathtt{v}_{j,\ell}$ (resp.\ $\mathtt{c}_{i,1},\dots,\mathtt{c}_{i,\ell}$).
The factor node $\mathtt{f}_{i}\in\mathtt{F}$ is replaced with $\ell+\ell_i$ nodes $\mathtt{f}_{i,1},\dots,\mathtt{f}_{i,\ell+\ell_i}$.
The memory of $\mathtt{c}_{i,t}$ (resp.\ $\mathtt{f}_{i,t}$) is set as the coefficient of $z^t$ in $w_i(z)$ (resp.\ $w_i'(z)$).
If the coefficient of $z^t$ in the label $\l_{i,j}$ is 1, $\mathtt{v}_{j,k}$ and $\mathtt{c}_{i,k+t}$ are connected by an edge for $k\in[1,\ell]$.
After constructing the bit-wise residual graph, the PA works on it.

In the case of ZD fountain codes, there is possibility that there exist some factor nodes of degree 1 in $\mathtt{G}_{\mathrm{b}}$ by the labels on the edges.
For example, a factor node in $\mathtt{G}$, which connects to the two edges with labels $1$ and $z$, gives the two factor node of degree 1 in $\mathtt{G}_{\mathrm{b}}$.
Hence, even if some precoded packets are not recovered by the packet-wise PA, there is possibility that the bit-wise PA recovers those precoded packets.
On the other hand, in the case of Raptor code, there exists no factor node of degree 1 in $\mathtt{G}_{\mathrm{b}}$ since all the edges are labeled by 1.
Hence, the bit-wise PA does not work for the Raptor codes.

\begin{algorithm}[tb]
 \begin{footnotesize}
  \caption{Packet-wise peeling algorithm \label{algc:1}}
  \begin{algorithmic}[1]
\REQUIRE Received packets ${\bs r}_1,\dots {\bs r}_{\tilde{k}}$ and packet-wise representation of factor graph $\mathtt{G}_{\mathrm{p}}$
\ENSURE Precoded packets ${\bs a}_1, \dots, {\bs a}_n$, residual graph $\mathtt{G}$, and values of memories $w_1(z),\dots,w_{m}(z), w_1'(z),\dots, w_{\tilde{k}}'(z)$
\STATE $\tau \ot 1$, $\mathtt{G}\ot \mathtt{G}_{\mathrm{p}}$
\STATE $\forall i\in[1,n]~~ {\bs a}_i = *^{\ell}$ \label{stp:init_a}
\STATE $\forall i\in[1,m]~~ w_i(z) \ot 0$ and $\forall i\in[1,k']~~ w_i(z) \ot r_{i}(z)$ 
\STATE Remove all the variable nodes in $\mathtt{V}_{\mathrm{r}}$ and those connecting edges from $\mathtt{G}$.

\FOR{$i\in[1,k']$} \label{stp:rec}
\IF{the degree of the $i$-th factor node $\mathtt{f}_i$ is 1}
\STATE Let $j$ be the position of variable node which is adjacent to $\mathtt{f}_i$.
\STATE $a_j(z)\ot \ell_{i,j}^{-1}s_i(z)$.
\STATE $\forall t\in\neib{vc}(j)~~ w_t(z)\ot w_t(z)+ a_j(z)$
\STATE $\forall t\in\neib{vf}(j)~~ w_t'(z)\ot w_t'(z) + \ell_{t,j} a_j(z)$
\STATE Remove the variable node $\mathtt{v}_{j}$ and its connecting edges from $\mathtt{G}$. 
\ENDIF
\ENDFOR

\FOR{$i\in[1,m]$} 
\IF{the degree of the $i$-th check node $\mathtt{c}_i$ is 1}
\STATE Let $j$ be the position of variable node which is adjacent to $\mathtt{f}_i$.
\STATE $a_j(z) \ot s_i(z)$.
\STATE $\forall t\in\neib{vc}(j)~~ w_t(z)\ot w_t(z)+ a_j(z)$ 
\STATE $\forall t\in\neib{vf}(j)~~ w_t'(z)\ot w_t'(z) + \ell_{t,j} a_j(z)$
\STATE Remove the variable node $\mathtt{v}_{j}$ and its connecting edges from $\mathtt{G}$. 
\ENDIF
\ENDFOR

\IF{ there exist check nodes or factor nodes of degree 1 in $\mathtt{G}$}
\STATE $\tau \ot \tau +1$ and go to Step \ref{stp:rec}
\ENDIF
  \end{algorithmic}
 \end{footnotesize}
\end{algorithm}

\begin{example}
Figure \ref{fig:fg1_pw} shows the residual graph $\mathtt{G}$ after the packet-wise PA for the ZDF code given in Fig.\ \ref{fig:foun}.
Figure \ref{fig:fg1_bw} shows the corresponding bit-wise representation, where $\ell = 4$.
In the similar way to PA over an LDPC code, 
decoding starts from the squares of degree 1.
In this case, the leftest and rightest black circles are decoded from the leftest and rightest white squares in Fig. \ref{fig:fg1_bw}.
The decoded circles and those connecting edges are removed from the bit-wise residual graph.
At the second iteration round, all the circles are decoded by in this procedure.

\begin{figure}[tb]
 \begin{center}
  \subfigure[Residual graph $\mathtt{G}$]{%
   \centering
   \setlength\unitlength{1pt}
    \begin{picture}(100,70)
     \put(15,0){\includegraphics[height=70pt]{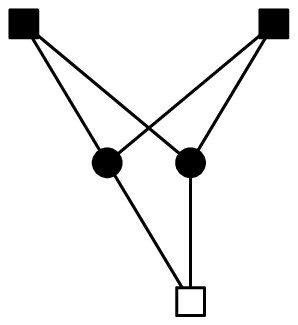}}
     \put( 40, 10){$1$}
     \put( 60, 10){$z$}
    \end{picture}
    \label{fig:fg1_pw}
  }
  \subfigure[Bit-wise representation of $\mathtt{G}$]{%
    \begin{picture}(100,70)
      \put(5,0){\includegraphics[height=70pt]{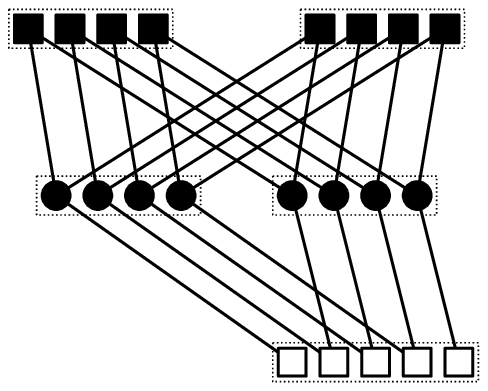}}
    \end{picture}
    \label{fig:fg1_bw}
  }
  \caption{ Residual graph and its bit-wise representation ($\ell=4$). \label{fig:bit_rep}}
 \end{center}
\end{figure}



\end{example}



\subsection{Related Work}
Qureshi {\it et al}.\ \cite{DBLP:journals/corr/abs-1305-0918} suggested a fountain coding system based on triangular coding.
This fountain coding system is an improvement of the LT code.
Note that, in this fountain coding system, the encoder chooses {\it distinct} $d$ shift amount $\delta_1,\dots,\delta_d$, namely,  $\delta_i \neq \delta_j$ for $i\neq j$.
The proposed encoding algorithm is regarded to as a generalization of the fountain code in \cite{DBLP:journals/corr/abs-1305-0918}, since the proposed encoding algorithm is an improvement of Raptor code and can choose $d$ shift amount $\delta_1,\dots,\delta_d$ with $\delta_i = \delta_j$ for $i\neq j$.

\section{Performance Evaluation \label{sec:pe}}
In this section, we evaluate the performance of the proposed fountain coding system.

\subsection{Overhead for the received bits}
For the proposed fountain code, the length of the output packets is slightly longer than that of the source packets.
Denote the length of the $i$-th received packet, as $\ell + \ell_i$, with $\ell_i \ge 0$.
Then, the total number of bits in the received packets is $\tilde{k}\ell + \sum_{i=1}^{\tilde{k}}\ell_i$, where $\tilde{k} = k(1+\alpha)$.
Hence, we need to consider not only the number of received packets $\tilde{k}$ but also the total number of the bits in the data section of the received packets\footnote{Simply, we refer to the total number of the bits in the data section of the received packets as the number of received bits.} $k\ell(1+\beta)$.
We refer to the value $\beta$ as the {\it overhead for the received bits}.
The value $\beta$ is given by
\begin{equation}
 \beta 
  = 
  \alpha 
 + \frac{\sum_{i=1}^{\tilde{k}} \ell_i}{k\ell}. \label{eq:beta1}
\end{equation}
Notice that for the Raptor codes, $\beta = \alpha$ holds since $\ell_i = 0$ for all $i$.
From the above equation, to calculate $\beta$, we need to evaluate $\ell_1,\ell_2,\dots, \ell_{\tilde{k}}$.

Let $L$ be a random variable which represents a length of a received packet.
For a given degree distribution $\Omega(x)$ and a given shift distribution $\Delta(x)$, the expectation of $L$ is given as the following proposition.
\begin{proposition} \label{lem:1} \upshape
For a given degree distribution $\Omega(x) = \sum_{i}\Omega_i x^i$ and a given shift distribution $\Delta(x) = \sum_{i=0}^{s_m}\Delta_i x^i$, the following holds:
\begin{align*}
 &\mathbb{E}[L]
  =
 \ell
  + s_m
  - \sum_{i=0}^{s_m-1} \Omega\bigl(\Delta_{[0,i]}\bigr)
  - \sum_{i=1}^{s_m  } \Omega\bigl(\Delta_{[i,s_m]}\bigr),
\end{align*}
where $\Delta_{[i,j]} := \sum_{t\in[i,j]}\Delta_t$.
\end{proposition}
The proof of this lemma is given in \ref{app:a}
When the shift distribution $\Delta(x)$ is a uniform distribution, we get the following corollary from Proposition \ref{lem:1}.
\begin{corollary} \label{cor:1} \upshape
When the shift distribution is a uniform distribution, i.e, $\Delta(x) = \sum_{i=0}^{s_m}x^{i}/(s_m+1)$, for a given degree distribution $\Omega(x)$, the following holds:
\begin{align*}
 &\mathbb{E}[L]
  =
 \ell + s_{m}- 2 \sum_{i=1}^{s_m} \Omega \left(\frac{i}{s_m+1}\right).
\end{align*}
\end{corollary}
{\it proof}:
Since the shift distribution is uniform distribution, $\Delta_{[0,i]} = (i+1)/(s_m+1)$ and $\Delta_{[i,s_m]} = (s_m-i+1)/(s_m+1)$ hold.
By substituting those equations into the equation in Proposition \ref{lem:1}, we get this corollary.
\hfill\QED

By using Proposition \ref{lem:1} and Eq.~\eqref{eq:beta1}, for a fixed $\alpha$, the expectation of the overhead for the received bits is
\begin{equation}
 \beta 
 =
 (1+\alpha) \frac{\mathbb{E}[L]}{\ell} - 1. \label{eq:exp_beta}
\end{equation}
Since $\ell \le \mathbb{E}[L]\le \ell + s_m$, $\beta \to \alpha$ as $\ell\to \infty$.

\subsection{Decoding Erasure Probability}
In this section, we compare the decoding erasure probability of the proposed fountain coding system with that of the Raptor coding system.

We denote the proposed fountain code with the precode $\mathcal{C}$, the degree distribution $\Omega(x)$ and the shift distribution $\Delta(x)$, as $\mathcal{F}(\mathcal{C},\Omega(x),\Delta(x))$.
Note that the Raptor code is a special case for the proposed fountain code with $\Delta(x) = 1$.
In other words, $\mathcal{F}(\mathcal{C},\Omega(x),1)$ represents the Raptor code with the precode $\mathcal{C}$ and the degree distribution $\Omega(x)$.
In this section, we will prove that the fountain code $\mathcal{F}(\mathcal{C},\Omega(x),\Delta(x))$ outperforms the Raptor code $\mathcal{F}(\mathcal{C}, \Omega(x),1)$ in terms of the decoding erasure probability.

To prove the above, we use the following lemma.
\begin{lemma} \label{lem:2} \upshape
Fix an unlabeled factor graph $\mathtt{G}$ in the packet-wise representation.
If decoding succeeds for the factor graph $\mathtt{G}$ in which all the edges are labeled by $1$, the decoding also succeeds for the factor graph $\mathtt{G}$ with arbitrary labeling.
\end{lemma}
{\it proof}:
We use the proof by contradiction.
We assume that decoding is failed for the factor graph $\mathtt{G}$ with some labeling.
Then, the factor graph $\mathtt{G}$ contains some stopping sets.
Thus, the decoding is also a failure for the factor graph $\mathtt{G}$ all the edges of which are labeled by $1$. 
\hfill\QED

This lemma shows that, for a fixed unlabeled factor graph, the decoding succeeds for the proposed fountain coding system if the decoding succeeds for the Raptor coding system.
From this lemma, we obtain the following theorem.
\begin{theorem} \label{the:1} \upshape
Let $\mathrm{P}(\alpha,\mathcal{C},\Omega(x),\Delta(x))$ be the decoding erasure probability for the fountain code $\mathcal{F}(\mathcal{C},\Omega(x),\Delta(x))$ at the overhead $\alpha$ for the received packets.
For arbitrary $\alpha, \mathcal{C}, \Omega(x),\Delta(x)$, the following holds:
\begin{equation*}
 \mathrm{P}(\alpha,\mathcal{C},\Omega(x), 1)
  \ge
 \mathrm{P}(\alpha,\mathcal{C},\Omega(x),\Delta(x)).
\end{equation*}
\end{theorem}

This theorem shows that the fountain code $\mathcal{F}(\mathcal{C},\Omega(x),\Delta(x))$ outperforms the Raptor code $\mathcal{F}(\mathcal{C}, \Omega(x),1)$ in terms of the decoding erasure probability.
As shown in \cite{Raptor}, Raptor codes have arbitrary small overhead $\alpha$ in large $k$.
Hence, the proposed fountain codes also have arbitrary small overhead $\alpha$ for the asymptotic case.

\subsection{Decoding Complexity \label{ssec:dec_comp}}
Recall that the proposed decoding algorithm works on the factor graph in the bit-wise representation in a similar way to the PA for the LDPC code over the BEC.
Hence the space complexity of the decoding algorithm is equal to the total number of factor nodes in inner code, precoded nodes and factor nodes in precode in the bit-wise representation.
Then, the space complexity of the decoding algorithm is $\ell(2n+\beta k)$.
Similarly, the decoding complexity of the decoding algorithm for the Raptor code is $\ell(2n+\alpha k)$, since $\beta = \alpha$ holds in the case of the Raptor code.

The number of iterations of the PA is upper bounded on the total number of check nodes and factor nodes in the factor graph.
For the Raptor coding system, since the decoding algorithm works on the factor graph in the packet-wise representation, the number of iterations is upper bounded on $\alpha k + n$.
On the other hand, for the proposed fountain coding system, the number of iterations is upper bounded on $\ell(\beta k + n)$ since the decoding algorithm works on the factor graph in the bit-wise representation.
This is the main drawback of the proposed fountain coding system.
However, the average number of decoding iterations is much smaller than the worst case.
In the following section, we will evaluate the average number of decoding iterations by numerical simulations.

\subsection{Numerical Example And Simulation Results \label{ssec:sr}}
This section shows a numerical example of the expected length of the received packets and simulation results which show the decoding erasure rates and the average number of decoding iterations.

As a precode, we employ (3,30)-regular LDPC codes with $(k,n) = (900,1000), (1800,2000), (3600,4000)$.
The degree distribution for the inner code is $\Omega(x) = 0.007969x +0.493570x^2 +0.166220x^3
+0.072646x^4 +0.032558x^5 +0.056058x^8 +0.037229x^9 +0.055590x^{19}
+0.025023x^{65} +0.003135x^{66}$ \cite{Raptor}.
The shift distribution is uniform distribution, i.e, $\Delta(x) = \sum_{i=0}^{s_m}x^i/(s_m+1)$.

\subsubsection{Length of Received Packets}

\begin{table}[tb]
 \caption{
The expected number of additional bits for the received packets, $\mathbb{E}[L]-\ell$, with $s_m=1,2,\dots, 6$. \label{tab:leng}}
 \begin{center}
  \begin{tabular}{|c|c|c|c|c|c|c|} \hline
   $s_m$ & 1 & 2 & 3 & 4 & 5 & 6 \\ \hline \hline
   $\mathbb{E}[L]-\ell$ & 0.6758 & 1.2412 & 1.7730 & 2.2896 & 2.7979 & 3.3011 \\ \hline 
  \end{tabular}
 \end{center}
\end{table}

In this section, we compute the expected number of additional bits in the received packets, $\mathbb{E}[L]-\ell$, from Corollary \ref{cor:1}.
Table \ref{tab:leng} gives $\mathbb{E}[L]-\ell$ with $s_m=1,2,\dots, 6$.
From Table \ref{tab:leng}, the expected number of additional bits in the received packets is monotonically increasing as $s_m$ increases.

\subsubsection{Decoding Erasure Rate}
\begin{figure}[tb]
 \begin{center}
  \subfigure[Decoding erasure rate for the overhead $\alpha$ \label{fig:DER1000_alpha}]{
   \begin{picture}(200,140)
    \put(0,0){\includegraphics[width=200pt]{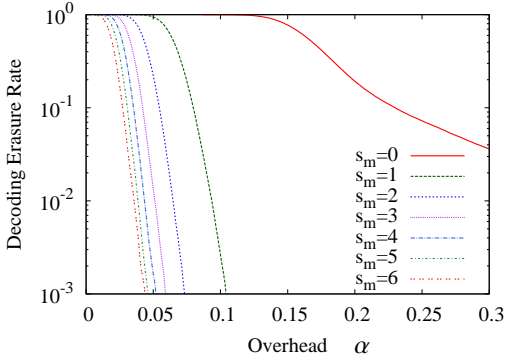}}
    \put(135,2.5){$\alpha$}
   \end{picture}
  }
  \subfigure[Decoding erasure rate for the overhead $\beta$ \label{fig:DER1000_beta}]{
   \begin{picture}(200,140)
    \put(0,0){\includegraphics[width=200pt]{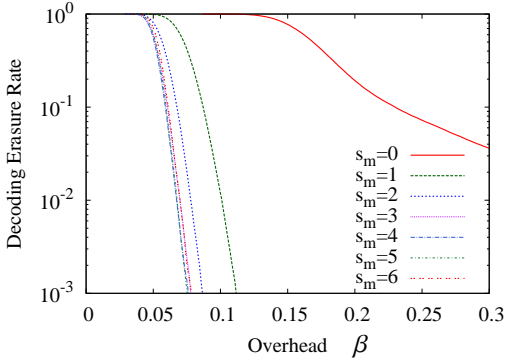}}
    \put(135,2.5){$\beta$}
   \end{picture}
  }
 \end{center}
 \caption{Decoding Erasure rate for the Raptor code ($s_m = 0$) and the proposed fountain code ($s_m = 1,2,\dots, 5$) with $k=900$ and $\ell = 100$.}
\end{figure}

The decoding erasure rate is the fraction of the trials in which some bits in the precoded packets are not recovered.
In this simulation, we examine the decoding erasure rates (DERs) of the Raptor code and the proposed fountain code.
Figures \ref{fig:DER1000_alpha}, \ref{fig:DER1000_beta} display the DERs for the Raptor code and the proposed fountain codes with $\ell = 100$.
The horizontal axis of Fig.~\ref{fig:DER1000_alpha} (resp.\ Fig.~\ref{fig:DER1000_beta}) represents the overhead $\alpha$ (resp.\ $\beta$).
The curve with $s_m=0$ gives the DER of the Raptor code.
The curves with $s_m=1,2,\dots, 6$ give the DERs of the proposed fountain codes with the maximum shift amount $s_m=1,2,\dots, 6$, respectively.

As shown in Fig.~\ref{fig:DER1000_alpha}, the DER is monotonically decreasing as the maximum shift amount $s_m$ increases.
Hence, the proposed fountain codes outperforms the Raptor code in terms of the DER for a fixed overhead $\alpha$.
In the case of overhead $\beta$, we see that from Fig.~\ref{fig:DER1000_beta} the DER does not monotonically decrease for the maximum shift amount $s_m$, 
i.e, the proposed fountain codes with $s_m=4$ or $s_m=5$ have good decoding performance.
This result is caused by the additional bits for the received packets, which increases as the maximum shift amount increases as shown in the previous section.

Moreover, from Figs.~\ref{fig:DER1000_alpha}, \ref{fig:DER1000_beta}, we see that the proposed fountain codes have smaller overheads $\alpha$ and $\beta$ than the Raptor code for a fixed DER. 
Since the space decoding complexity depends on the overhead $\beta$ as discussed in Section \ref{ssec:dec_comp}, the proposed fountain codes have smaller space decoding complexity than the Raptor code for a fixed DER.

\begin{figure}[tb]
 \begin{center}
  \subfigure[Decoding erasure rate for the overhead $\alpha$ \label{fig:DER_leng_alpha}]{
   \begin{picture}(200,140)
    \put(0,0){\includegraphics[width=200pt]{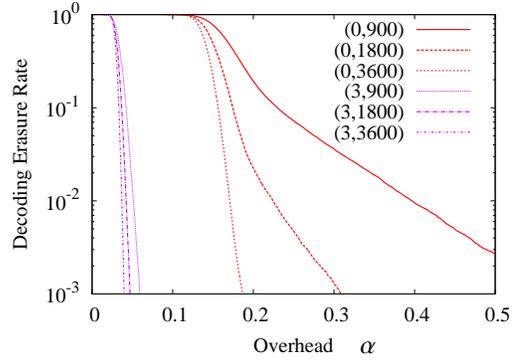}}
    \put(135,2.5){$\alpha$}
   \end{picture}
  }
  \subfigure[Decoding erasure rate for the overhead $\beta$ \label{fig:DER_leng_beta}]{
   \begin{picture}(200,140)
    \put(0,0){\includegraphics[width=200pt]{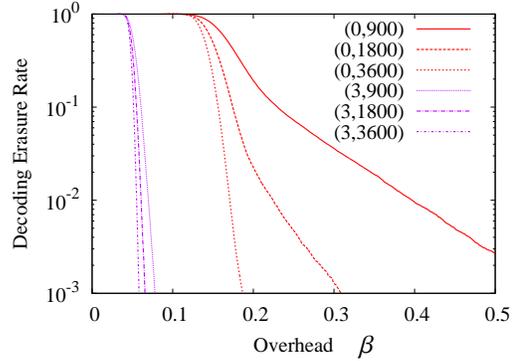}}
    \put(135,2.5){$\beta$}
   \end{picture}
  }
 \end{center}
 \caption{Decoding Erasure rate for the Raptor code ($s_m = 0$) and the proposed fountain code ($s_m = 3$) with $k=900,1800,3600$ and $\ell = 100$, where the labels of curves represent $(s_m,k)$.}
\end{figure}

Figures \ref{fig:DER_leng_alpha}, \ref{fig:DER_leng_beta} compare the DERs with various numbers of source packets, i.e, $k=900,1800,3600$.
The horizontal axis of Fig.~\ref{fig:DER_leng_alpha} (resp.\ Fig.~\ref{fig:DER_leng_beta}) represents the overhead $\alpha$ (resp.\ $\beta$).
The curves labeled with $(0,900)$, $(0,1800)$, and $(0,3600)$ give the DERs of the Raptor code with $k=900,1800$ and $k=3600$, respectively.
Similarly, the curves labeled with $(3,900)$, $(3,1800)$ and $(3,3600)$ represent the DERs of the proposed fountain code for $s_m = 3$ with $k=900,1800$ and $k=3600$, respectively.
From both Fig.~\ref{fig:DER_leng_alpha} and Fig.~\ref{fig:DER_leng_beta}, the DERs decrease as the number of source packets increases.
Moreover, the proposed fountain code outperforms the Raptor code in terms of the DER for every $k$.
In addition, we are able to expect that the decoding performance for $k\to \infty$ depends on the maximum shift amount $s_m$.
We will analyze the asymptotic decoding performance in Section \ref{sec:asympto}.

\subsubsection{Average Number of Iterations}
\begin{figure}[tb]
 \begin{center}
  \subfigure[$\ell = 100$. \label{fig:iter_num_100}]{
   \begin{picture}(200,140)
    \put(0,0){\includegraphics[width=200pt]{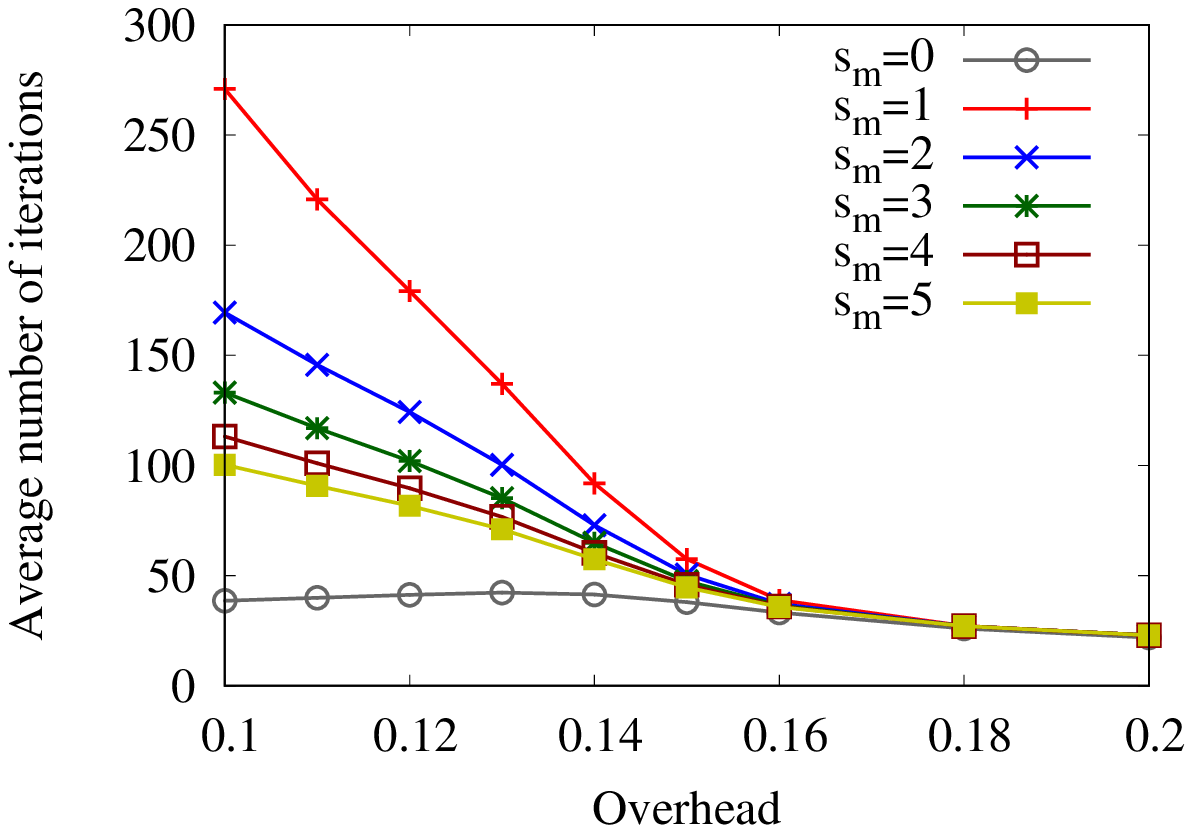}}
    \put(135,2.5){$\alpha$}
   \end{picture}
 }
 \subfigure[$\ell = 1000$. \label{fig:iter_num_1000}]{
   \begin{picture}(200,140)
    \put(0,0){\includegraphics[width=200pt]{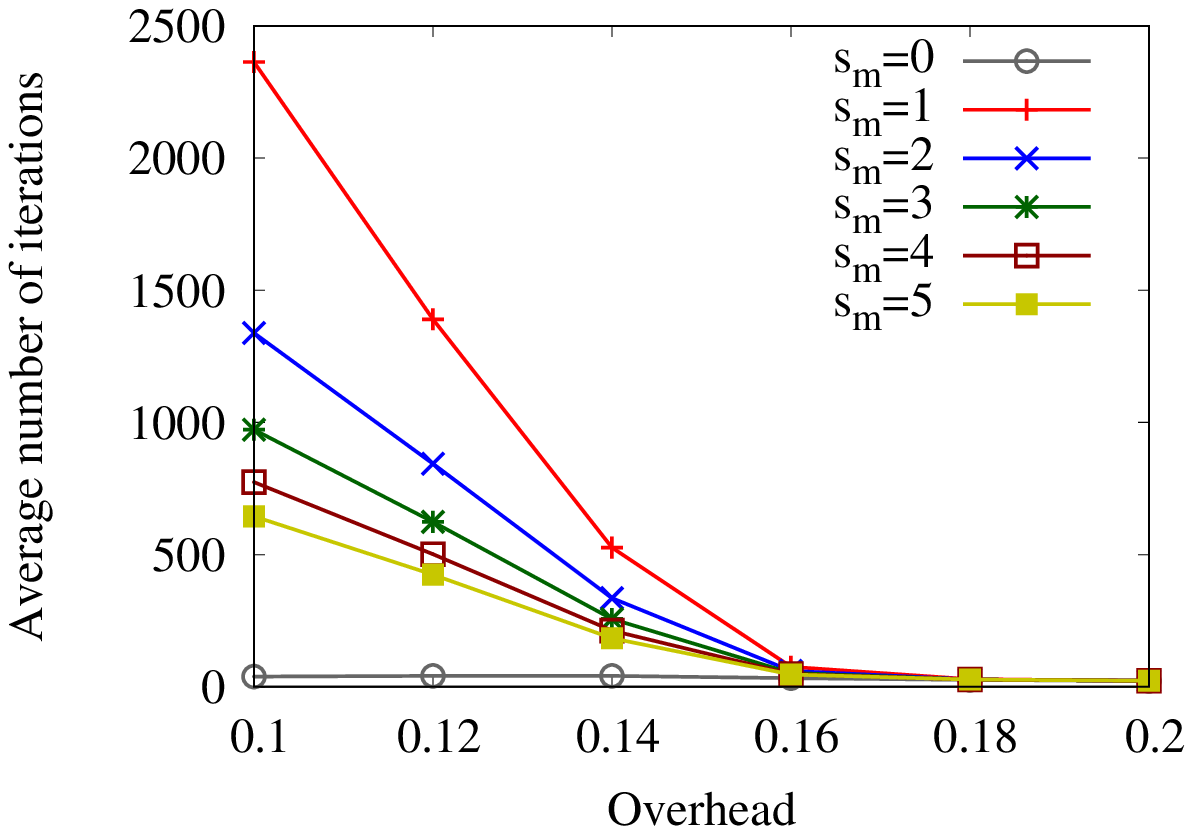}}
    \put(135,2.5){$\alpha$}
   \end{picture}
 }
 \end{center}
 \caption{The average number of decoding iterations for the proposed fountain codes with $s_m = 1,2,\dots,5$ and for the Raptor code ($s_m=0$) \label{fig:iter_num}}
\end{figure}

The number of iterations is measured by the sum of iterations in packet-wise PA and bit-wise PA.
Figure \ref{fig:iter_num} depicts the average number of decoding iterations for the proposed fountain codes with $s_m = 1,2,\dots, 5$ and for the Raptor code ($s_m=0$).
We set the length of source packets $\ell=100$ (resp.\ $\ell=1000$) and the number of source packets $k=3600$ in Fig.~\ref{fig:iter_num_100} (resp.\ Fig.~\ref{fig:iter_num_1000}).
In this simulation, the decoding of the proposed fountain codes successes for all the trials.
On the other hands, the decoding of the Raptor codes does not success on some trials.
From Fig.~\ref{fig:iter_num}, the proposed fountain codes need more decoding iterations than the Raptor codes.
The gap of average number of decoding iterations between Raptor codes and the proposed fountain code gives the number of bit-wise decoding iterations.
Hence this gap represents the additional decoding iterations for successful decoding.

From Fig.~\ref{fig:iter_num}, the average number of iterations decreases as the overhead $\alpha$ increases.
The reason is explained as the following:
For a fixed un-labeled factor graph, if decoding of the Raptor code succeeds, then the proposed fountain code is decoded only by packet-wise PA.
Namely, the proposed fountain code is decoded in the same number of iterations for the Raptor code.
Hence, the number of iterations converges to the same value as $\alpha$ increases.
Moreover, from Fig.~\ref{fig:iter_num}, the average number of iterations decreases as the maximum shift amount $s_m$ increases for $s_m \ge 1$.
This implies that the bit-wise PA recovers more bits in each decoding iteration as $s_m$ increases.

By comparing Figs.~\ref{fig:iter_num_100} and \ref{fig:iter_num_1000}, 
we see that the number of iterations increases as the length of packet $\ell$ increases.
The average number of bit-wise PA iterations for $\ell = 1000$ is about 10 times larger than that for $\ell=100$.
Namely, the average number of bit-wise PA iterations are proportional to $\ell$.

\subsubsection{Summary of Simulation Results}
From the above simulation results, for a fixed DER, the proposed fountain code has smaller overheads $\alpha,\beta$ than the Raptor code.
This implies that the proposed fountain code has smaller space decoding complexity than the Raptor code from Section \ref{ssec:dec_comp}.
However, the proposed fountain code has a greater average number of decoding iterations is greater than the Raptor code.
Namely, the proposed fountain code is inferior to the Raptor code in terms of decoding latency.

For the proposed fountain code, as the maximum shift amount $s_m$ increases, the DER and average number of decoding iterations monotonically decrease for a fixed overhead $\alpha$.
On the other hand, for a fixed $\beta$, the DER does not monotonically decrease as the maximum shift amount $s_m$ increases.
In other words, there exist finite optimal $s_m$ for the DER in regard to the overhead for the received bits $\beta$.

\section{Asymptotic Analysis \label{sec:asympto}}
In this section, we will analyze the decoding performance for the ZD fountain code in the limit of a large number of source packets by using density evolution.
To introduce the density evolution of the ZD fountain code, firstly we will give a belief propagation (BP) in Section \ref{ssec:mpa}.
In Section \ref{ssec:DE}, we will give the density evolution equations.
Section \ref{ssec:DE_NE} gives a numerical example for the asymptotic overheads $\alpha^*$ and $\beta^*$ by using density evolution.

\subsection{Belief Propagation \label{ssec:mpa}}
This section gives the BP algorithm for the ZD fountain codes.
This BP algorithm can be straightforwardly extended from the BP algorithm for the LDPC codes through the BEC.
In the BP, decoding proceeds by sending messages along the edges in the factor graph.
Each message in the decoder is given by a vector of length $\ell$.
We denote the $t$-th entry of the message ${\bs \mu}$, by $\mu[t]$.
Each entry of the messages is a member of $\{0,1,*\}$, where $*$ represents an erasure.

Initially, for all $i\in[1,\tilde{k}]$, the $i$-th factor node of the inner code stores the received packet ${\bs r}_i$ in the memory.
The decoder peels off the variable nodes representing received packets and those connecting edges from the factor graph.

\begin{figure}[tb]
 \begin{center}
  \begin{picture}(70,70)
   \put(0,0){\includegraphics[height=70pt]{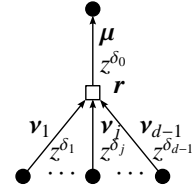}}
   \put(13,10){$z^{\delta_1}$}
   \put(32,10){$z^{\delta_j}$}
   \put(52,10){$z^{\delta_{d-1}}$}
   \put(32,43){$z^{\delta_0}$}
   \put( 5,20){${\bs \nu}_1$}
   \put(32,20){${\bs \nu}_j$}
   \put(47,20){${\bs \nu}_{d-1}$}
   \put(32,55){${\bs \mu}$}
   \put(37,35){${\bs r}$}
   \put(12,2){$\cdots$}
   \put(38,2){$\cdots$}
  \end{picture}
  \caption{Message flow in a factor node of degree $d$ \label{fig:fnode}}
 \end{center}
\end{figure}

In the iteration step, each node generates the outgoing messages from the incoming messages.
In the case of a variable node of degree $d$, the outgoing message ${\bs \nu}$ is generated from the $d-1$ incoming messages ${\bs \mu}_1, \dots, {\bs \mu}_{d-1}$ as the following rule:
\begin{equation}
 \nu[t]
  =
 \begin{cases}
  0, & \exists j \text{~s.t.~} \mu_j[t] = 0, \\
  1, & \exists j \text{~s.t.~} \mu_j[t] = 1, \\
  *, & \forall j ~ \mu_j[t] = *.
 \end{cases}
\end{equation}
Note that there is no possibility for existing $i,j$ such that $\mu_i[t] = 0$ and $\mu_j[t] = 1$ because of erasure decoding.
For a check node of degree $d$ on the precode, the outgoing message ${\bs \mu}$ is generated from the $d-1$ incoming message ${\bs \nu}_1, \dots, {\bs \nu}_{d-1}$ as the following rule:
\begin{equation}
 \mu[t]
  =
 \begin{cases}
  \sum_{j=1}^{d-1} \nu_{j}[t], & \forall j ~ \nu_j[t] \neq *, \\
  *, & \text{Otherwise}. 
 \end{cases}
\end{equation}
Finally, we consider the update rule for the factor node of degree $d$.
Figure \ref{fig:fnode} depicts a message flow of the factor node.
In words, the edge sending outgoing message ${\bs \mu}$ is labeled by $z^{\delta_0}$ and
the edge sending incoming message ${\bs \nu}_j$ is labeled by $z^{\delta_j}$.
Assume that the check node stores ${\bs r}$ in its memory.
Recall that the factor node on the inner code represents a constraint given in Eq.~\eqref{eq:constraint}.
Denote $t_j' := t + \delta_0 - \delta_j$.
The constraint given in Eq.~\eqref{eq:constraint} gives the following update rule:
\begin{align}
 &\mu[t]
  =
 \begin{cases}
  r_{t + \delta_0} + \sum_{j=1}^{d-1} \nu_{j}[t'_{j}],
  & \forall j~ \nu_j[t'_{j}] \neq *, \\
  *, 
  & \text{Otherwise},
  \\
 \end{cases} 
\end{align}
where $\nu_j[t] = 0$ if $t \le 0$ or $t > \ell$.

In the marginal step, the decoder decides the decoding output $(\hat{x}_1,\dots,\hat{x}_{\ell})$ at each decoding round.
Each variable node of degree $d$ decides the decoding output from all the $d$ incoming messages, ${\bs \mu}_1,\dots, {\bs \mu}_{d}$ as the following rule:
\begin{align}
 \hat{x}_{t}
  =
 \begin{cases}
  0, & \exists j \text{~s.t.~} \mu_j[t] = 0, \\
  1, & \exists j \text{~s.t.~} \mu_j[t] = 1, \\
  *, & \forall j ~ \mu_j[t] = *.
 \end{cases}
\end{align}

\subsection{Density Evolution \label{ssec:DE}}
The system of density evolution equations for the erasure channels tracks the erasure probabilities of the messages at each iteration of the BP.
For the ZD fountain codes, since each message is given in a vector of length $\ell$, the erasure probabilities of messages are represented in vectors of length $\ell$.
Let ${\bs x}_1^{(\tau)} = (x_{1,i}^{(\tau)})_{i=1}^{\ell}$ be the erasure probability of the message from the variable nodes $\mathtt{V}_{\rm p}$ to the check nodes $\mathtt{C}$ at the $\tau$-th iteration, where $x_{1,i}^{(\tau)}$ gives the erasure probability of the $i$-th entry in the message.
Let ${\bs x}_2^{(\tau)} = (x_{2,i}^{(\tau)})_{i=1}^{\ell}$ be the erasure probability of the message from the variable nodes $\mathtt{V}_{\rm p}$ to the factor nodes $\mathtt{F}$ at the $\tau$-th iteration.
Similarly, we denote the erasure probability of the message from the check nodes $\mathtt{C}$ (resp.\ from the factor nodes $\mathtt{F}$) to the variable nodes $\mathtt{V}_{\rm p}$ at the $\tau$-th iteration, by ${\bs y}_1^{(\tau)}$ (resp.\ ${\bs y}_2^{(\tau)}$).

In this section, we assume that the precode is chosen from the irregular LDPC ensemble with node degree distribution $\Lambda(x)=\sum_d \Lambda_d x^d$ and $P(x) = \sum_d P_d x^d$, where $\Lambda_d$ (resp.\ $P_d$) is the fraction of the variable node (resp.\ the check node) of degree $d$.
Then, the edge degree distribution $\lambda(x)=\sum_d \lambda_d x^{d-1}$ and $\rho(x)=\sum_d \rho_d x^{d-1}$ of this LDPC code is given by $\lambda(x) = \Lambda'(x)/\Lambda'(1)$ and $\rho(x) = P'(x)/P'(1)$, respectively.
We denote the rate of the LDPC code, by $R$, i.e, $R = 1 - \Lambda'(1)/P'(1)$.

Next, we consider the degree distributions of the inner code side.
In a factor node $\mathtt{f}\in\mathtt{F}$, we refer to the number of edges connecting to the variable nodes $\mathtt{V}_{\rm p}$ as the degree of $\mathtt{f}$.
From the encoding algorithm, $\Omega(x)$ gives the node degree distribution of the factor nodes.
The edge degree distribution for the factor node, denoted by $\omega(x) = \sum_{d}\omega_d x^{d-1}$, is given by $\omega(x) = \Omega'(x)/\Omega'(1)$
Let $I_d$ denote the expected fraction of variable node of degree $d$ in the inner code side.
As shown in \cite[Section VI]{Raptor}, $I_d$ is derived as
$I_d
  =
 \binom{\tilde{k}}{d} \left( \bar{\Omega}/n \right)^{d}
 \left(1- \bar{\Omega}/n \right)^{\tilde{k}-d},
$
where $\bar{\Omega} = \Omega'(1)$.
Hence, the node degree distribution of variable nodes, defined by $I(x) := \sum_d I_d x^d$, is 
\begin{align*}
 I(x)
 =&
 \left( 1 + \bar{\Omega}(x-1)/n \right)^{\tilde{k}} \\
 \to&
 \exp \left[ \bar{\Omega} R (1+\alpha) (x-1) \right]\quad (k\to \infty)
\end{align*}
Then, the edge degree distribution $\iota(x) = \sum_{d} \iota_dx^{d-1}$ is given by $\iota(x) = I'(x)/I'(1)$. 
For $k\to\infty$, $\iota(x) = I(x)$ holds.

From the above, each variable node has two type degrees in the precode side and inner code side.
The degree of a variable node is denoted by $(d_1,d_2)$ if the variable node is of degree $d_1$ in the precode side and of degree $d_2$ in the inner code side.
Since the degrees of the precode side and inner code side are decided independently each other, the fraction of variable nodes of degree $(d_1,d_2)$ is $\Lambda_{d_1}I_{d_2}$.

In this section, we denote the labels of the edges by the un-normalized shift amount.
In other words, the edge $(\mathtt{f}_i, \mathtt{v}_j)$ is labeled by $z^{\tilde{\delta}_{i,j}}$ instead of $z^{\delta_{i,j}}=z^{\tilde{\delta}_{i,j}-\min_j \tilde{\delta}_{i,j}}$.
Even if we label the un-normalized shift amount on the edges, the decoding result is equivalent to the normalized shift case.

Initially, all the messages from the variable node are erased.
Hence, for all $i\in[1,\ell]$, $x^{(0)}_{1,i} = x^{(0)}_{2,i} = 1$ holds.

Next, we derive the density evolution equations from the iteration step of the BP.
Firstly, we derive the density evolution equation for ${\bs y}_1^{(\tau)}$ from the decoding process in the check nodes.
The probability that an edge $\mathtt{e}$ connects to a check node of degree $d$ is $\rho_d$.
The erasure probability of the $i$-th entry of the message to a variable node in the chosen edge $\mathtt{e}$ at the $\tau$-th iteration is $1 - \{1-(x_{1,i}^{(\tau-1)})\}^{d-1}$.
Hence, we get
\begin{equation}
 y_{1,i}^{(\tau)} = 1- \rho\Bigl( 1- x_{1,i}^{(\tau-1)}\Bigr). \label{eq:DE_y1}
\end{equation}
Secondly, we derive the density evolution equation for ${\bs x}_1^{(\tau)}$ and ${\bs x}_2^{(\tau)}$ from the decoding process in the variable nodes.
The probability that an edge $\mathtt{e}$ in the precode side connects to a variable node of degree $(d_1,d_2)$ is $\lambda_{d_1}I_{d_2}$.
The erasure probability of the $i$-th entry of the message to a check node in the chosen edge $\mathtt{e}$ at the $\tau$-th iteration is $(y_{1,i}^{(\tau)})^{d_1-1}(y_{2,i}^{(\tau)})^{d_2}$.
Hence, we have
\begin{align}
 x_{1,i}^{(\tau)} 
  &= {\textstyle \sum_{d_1}\sum_{d_2}} \lambda_{d_1}I_{d_2}
  \Bigl(y_{1,i}^{(\tau)}\Bigr)^{d_1-1}\Bigl(y_{2,i}^{(\tau)}\Bigr)^{d_2} \notag \\
  &= \lambda\Bigl(y_{1,i}^{(\tau)}\Bigr)I\Bigl(y_{2,i}^{(\tau)}\Bigr) \label{eq:DE_x1}
\end{align}
Similarly, we get
\begin{align}
 x_{2,i}^{(\tau)} 
  = \Lambda\Bigl(y_{1,i}^{(\tau)}\Bigr)\iota\Bigl(y_{2,i}^{(\tau)}\Bigr) \label{eq:DE_x2}
\end{align}
Thirdly, we derive the density evolution equation for ${\bs y}_2^{(\tau)}$ from the decoding process in the factor nodes.
Let $q_{d,{\bs s}}$ denote the probability that a chosen edge $\mathtt{e}$ satisfies the following conditions:
the edge $\mathtt{e}$ is labeled by $z^{s_0}$, the connecting factor node $\mathtt{f}$ is of degree $d$, and the other $d-1$ edges of the factor node $\mathtt{f}$ are labeled with $z^{s_1},z^{s_2},\dots, z^{s_{d-1}}$. 
Then, the probability $q_{d,{\bs s}}$ is
\begin{equation}
 q_{d,{\bs s}}
 = 
 \omega_d \Delta_{s_0} {\textstyle \prod_{j=1}^{d-1}}\Delta_{s_j}. \notag
\end{equation}
The erasure probability of the $i$-th entry of the message to a variable node in the edge $\mathtt{e}$ at the $\tau$-th iteration is
\begin{equation}
 1-{\textstyle \prod_{j=1}^{d-1}}\Bigl(1-x_{2,i+s_0-s_j}^{(\tau-1)}\Bigr) \notag
\end{equation}
where $x_{2,r}^{(\tau)}=0$ if $r\le 0$ or $\ell < r$.
Hence, we have
\begin{align}
 y_{2,i}^{(\tau)}
  &= 
 {\textstyle \sum_{d}\sum_{s_0,s_1,\dots,s_{d-1}}} q_{d,{\bs s}} 
 \Bigl\{ 1-{\textstyle \prod_{j=1}^{d-1} } \Bigl(1-x_{2,i+s_0-s_j}^{(\tau-1)}\Bigr) \Bigr\} 
    \notag \\
  &=
 1 - 
{\textstyle 
\sum_{s_0}\Delta_{s_0} \sum_{d}\omega_d  \prod_{j=1}^{d-1}\sum_{s_j}\Delta_{s_j}\Bigl(1-x_{2,i+s_0-s_j}^{(\tau-1)}\Bigr) .}
 \notag
\end{align}
To simplify the notation, we denote 
\begin{equation*}
 \hat{x}_{2,i}^{(\tau)} 
 := 
 {\textstyle \sum_{s=0}^{s_m} } \Delta_{s}x_{2,i-s}^{(\tau)}.
\end{equation*}
Then the above equation can be transformed as follows:
\begin{align}
 y_{2,i}^{(\tau)} 
  &=
{\textstyle
 1 -  \sum_{s_0=0}^{s_m}\Delta_{s_0} \sum_{d}\omega_d \prod_{j=1}^{d-1}\Bigl(1-\hat{x}_{2,i+s_0}^{(\tau-1)}\Bigr) }\notag \\
  &=
{\textstyle
 1 -  \sum_{s_0=0}^{s_m}\Delta_{s_0}\sum_{d} \omega_d  \Bigl(1-\hat{x}_{2,i+s_0}^{(\tau-1)}\Bigr)^{d-1}}
 \notag \\
  &=
 1 -  {\textstyle \sum_{s=0}^{s_m} }\Delta_{s}\omega\Bigl(1-\hat{x}_{2,i+s}^{(\tau-1)}\Bigr) . \label{eq:DE_y2}
\end{align}

Finally, we derive that the $i$-th bit of the precode packets is erased at the $\tau$-th iteration, denoted by $Q_i^{(\tau)}$, from the marginal step of the BP.
The probability that a chosen variable node has degree $(d_1,d_2)$ is $\Lambda_{d_1}I_{d_2}$.
The $i$-th bit of this variable node is erased if $i$-th entries of all the incoming messages are erased.
Hence, we have
\begin{equation}
 Q_i^{(\tau)} = \Lambda\Bigl(y_{1,i}^{(\tau)}\Bigr)I\Bigl(y_{2,i}^{(\tau)}\Bigr) \label{eq:DE_Q}
\end{equation}

From the above, we have the system of density evolution equations as Eqs.~\eqref{eq:DE_y1}, \eqref{eq:DE_x1}, \eqref{eq:DE_x2}, \eqref{eq:DE_y2} and \eqref{eq:DE_Q}.
The decoding successfully stops if $Q_i^{(\tau)} = 0$ for all $i\in[1,\ell]$ at an iteration round $\tau$.
The overhead $\alpha^*$ in the case of $k\to\infty$ is obtained from the following equation:
\begin{equation}
 \alpha^* = \min\bigl\{ \alpha \mid \exists \tau \forall i\in[1,\ell] ~ Q_i^{(\tau)} = 0\bigr\}.
\end{equation}
The overhead $\beta^*$ is determined from $\alpha^*$ and Eq.~\eqref{eq:exp_beta} as follows:
\begin{equation}
 \beta^* = (1+\alpha^*) \frac{\mathbb{E}[L]}{\ell} -1.
\end{equation}
For a fixed maximum shift amount $s_m$, $\beta^* \to \alpha^*$ as $\ell\to \infty$.

\subsection{Numerical Example \label{ssec:DE_NE}}

\begin{table}[t]
 \begin{center}
  \caption{Overhead $\alpha^{*}$ for $k\to\infty$ \label{tab:th_al}}
  \begin{tabular}{|ll|cccccc|} \hline
         &       & \multicolumn{6}{c|}{$s_m$} \\
         &       & $0$  & $1$  & $2$  & $3$ & $4$ & $5$ \\ \hline
         & $16$  & 0.1282 & 0.0561 & 0.0338 & 0.0171 & -0.0011 & -0.0244 \\
         & $32$  & 0.1282 & 0.0563 & 0.0365 & 0.0265 & 0.0205  &  0.0156 \\
  $\ell$ & $64$  & 0.1282 & 0.0563 & 0.0365 & 0.0269 & 0.0219  &  0.0189 \\
         & $128$ & 0.1282 & 0.0563 & 0.0365 & 0.0269 & 0.0220  &  0.0190 \\
         & $256$ & 0.1282 & 0.0563 & 0.0365 & 0.0269 & 0.0220  &  0.0190 \\ \hline
  \end{tabular}
 \end{center}
\end{table}

\begin{figure}
 \begin{center}
  \includegraphics[width=240pt]{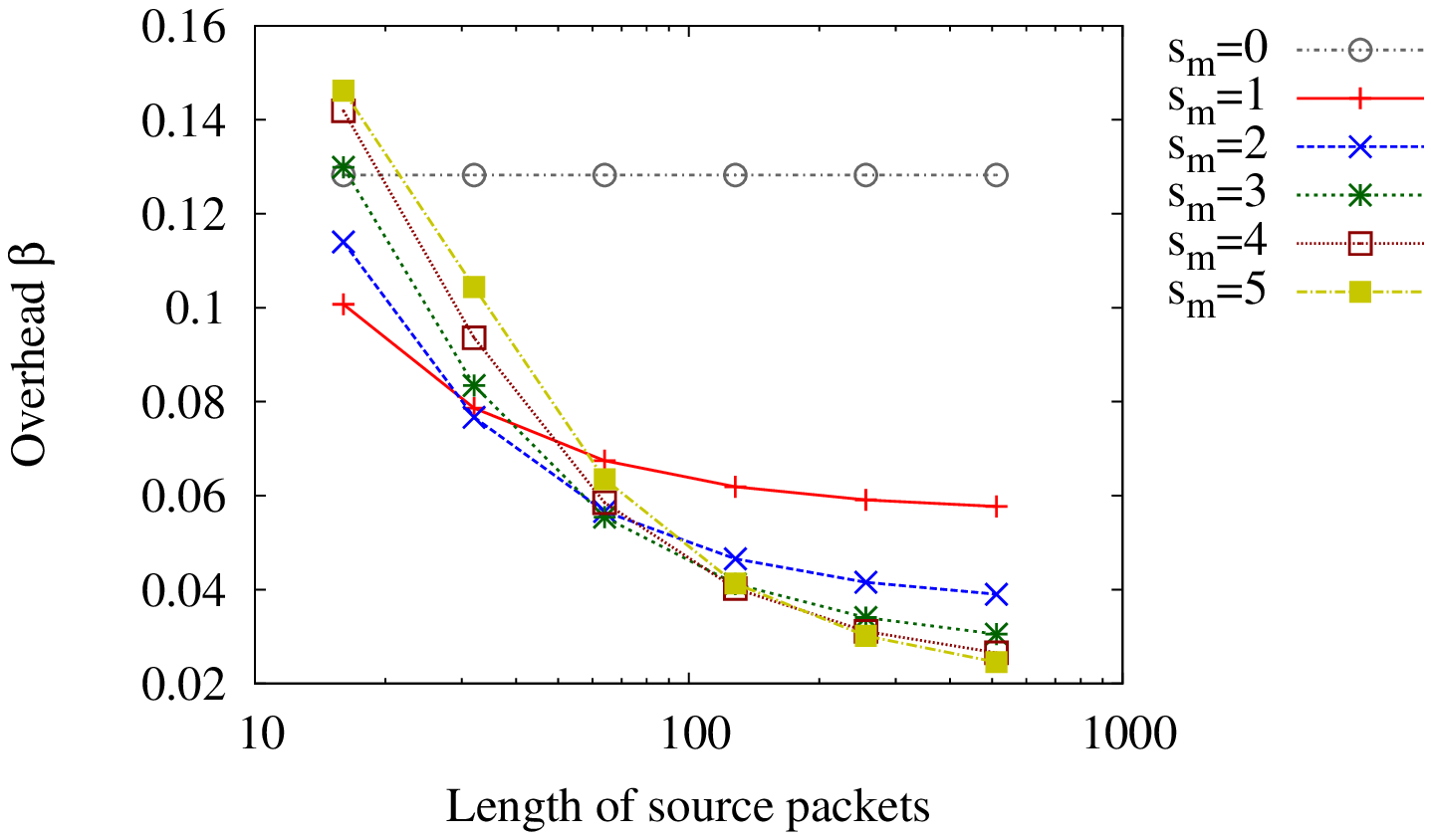}
  \caption{Overhead $\beta^{*}$ for $k\to\infty$ \label{fig:th_be}}
 \end{center}
\end{figure}

In this section, we evaluate the asymptotic overhead $\alpha^*$ and $\beta^*$ by a numerical example.
In this section, we use the same precode and degree distribution $\Omega(x)$ given in Section \ref{ssec:sr}.

Table \ref{tab:th_al} displays the overhead $\alpha^*$ for the proposed fountain codes.
From Table \ref{tab:th_al}, we see that the overhead $\alpha^*$ is monotonically decreasing as the maximum size of the shift amount $s_m$ increases at each packet length $\ell$.
Moreover, it is shown that the overhead $\alpha^*$ converges a certain value as $\ell$ increases.

To compare with the results in Fig.~\ref{fig:DER_leng_alpha}, we consider the case of $\ell = 100$.
As the results in Table \ref{tab:th_al}, the overhead $\alpha^*$ is 0.1282 (resp.\ 0.269) for $s_m = 0$ (resp.\ $s_m=3$).
From Fig.~\ref{fig:DER_leng_alpha}, we confirm that the decoding erasure rates are steeply down around those values.

Figure \ref{fig:th_be} shows the overhead $\beta^*$ for the proposed fountain codes.
From Fig.~\ref{fig:th_be}, we see that the overhead $\beta^*$ is monotonically decreasing as the length of source packets $\ell$ increases for each $s_m$.
In addition, we know that the optimum $s_m$ in terms of overhead $\beta^*$ depends on the length of source packets $\ell$.

We find that the fountain codes with $s_m = 4,5$ achieve $\alpha^* < 0$ for $\ell = 16$ from Table \ref{tab:th_al}.
This means that the $k$ source packets are decoded from the $\tilde{k}$ received packets where $\tilde{k}< k$.
However, even in those cases, we see that the number of received bits is greater than the source bits since the $\beta^{*}>0$ from Fig.~\ref{fig:th_be}.

\section{Conclusion \label{sec:conc}}
In this paper, we have proposed a fountain coding system based on ZD coding.
We have shown that the space complexity of the decoding algorithm for the proposed fountain coding system and the Raptor coding system depends on the received bits.
We have proved that the decoding erasure probability of the proposed fountain coding system is lower than that for the Raptor coding system for a fixed precode, degree distribution and overhead $\alpha$.
Moreover, we have shown that the proposed fountain coding system outperforms the Raptor coding system in terms of the overhead for the received packets and the received bits by simulation results.
Furthermore, we have derived the system of density evolution equations and have evaluated the asymptotic overheads $\alpha^*, \beta^*$ for a proposed fountain code.

\section*{Acknowledgment}
This work was partially supported by JSPS KAKENHI Grant Number 25889061.

\appendix
\section{Proof of Proposition \ref{lem:1} \label{app:a}}

To prove Proposition \ref{lem:1}, we use the following lemma.
\begin{lemma} \label{lem:min_max}
Suppose that $X_1, X_2, \dots, X_d \in [1,s_m]$ are i.i.d.\ discrete random variables.
Define $p_{[i,j]} := \Pr (i \le X_t \le j)$ for $t\in[1,d]$.
Denote ${\bf X} = \{X_1,X_2, \dots, X_d\}$.
Then the following equation holds:
\begin{align}
 &\Pr (\min{\bf X} = i, \max{\bf X} = j) \notag \\
 &\quad=
 p_{[i,j]}^d - p_{[i+1,j]}^d - p_{[i,j-1]}^d + p_{[i+1,j-1]}^d.
\end{align}
\end{lemma}
{\it proof}:
Define a random variable $N_t$ as $N_t := |\{ i\in[1,d] \mid X_i = t \}|$.
In words, $N_t$ represents the number of random variables which equal to $t$.
Assume that $\min {\bf X} = i$ and $\max {\bf X} = j$.
Notice that $N_i \ge 1$, $N_j \ge 1$ and $\sum_{t=i}^j N_i = d$.
Then, we have
\begin{align*}
 &\Pr(\min {\bf X} = i, \max {\bf X} = j, N_i =n_i, N_j = n_j) \\
  &\quad=
 \binom{k}{n_i, n_j, d-n_i-n_j} p_{[i,i]}^{n_i} p_{[j,j]}^{n_j} p_{[i+1,j-1]}^{d-n_i-n_j},
\end{align*}
where $\binom{d}{a,b,d-a-b}$ represents the multinomial coefficient, i.e,
$\binom{d}{a,b,d-a-b} = \frac{d!}{a! b! (d-a-b)!}$.
Thus, by using the above equation, we have
\begin{align*}
&\quad\Pr (\min {\bf X} = i, \max {\bf X} = j) \notag \\
 &= 
 \sum_{n_i = 1}^{d-1} \sum_{n_j = 1}^{d-n_i}
 \Pr(\min {\bf X} = i, \max{\bf X} = j, N_i =n_i, N_j = n_j) \\
 &= 
p_{[i,j]}^d - p_{[i+1,j]}^d - p_{[i,j-1]}^d + p_{[i+1,j-1]}^d.
\end{align*}
This concludes the proof.
\hfill\QED

Using the above lemma, we will prove Proposition \ref{lem:1}.

\noindent{\it Proof of Proposition \ref{lem:1}:}
Let $D$ be a random variable representing the degree of a chosen received node.
Then, the expectation of $L$ is derived as:
\begin{align}
 \mathbb{E}[L]
  &=
 {\textstyle \sum_{d}\sum_{j=0}^{s_m} }
 (\ell + j) \Pr [ L = \ell+j, D = d] \notag \\
  &=
 \ell + {\textstyle \sum_{d}} \Pr[D = d] {\textstyle \sum_{j=0}^{s_m}} j \Pr [ L = \ell+j \mid D = d]. \label{eq:app_a1}
\end{align}
Notice that $\Pr [D = d] = \Omega_d$.
Hence, we will consider $\sum_{j=0}^{s_m} j \Pr[L = \ell+j \mid D = d]$ to derive $\mathbb{E}[L]$.
From Eq.~\eqref{eq:leng}, the length of the received packet of degree $d$ with the shift amounts $(\tilde{\delta}_1,\dots, \tilde{\delta}_d)$ is
$\ell + \max_i \tilde{\delta}_i - \min_i \tilde{\delta}_i$.
Hence, we get
\begin{align*}
 &\Pr[ L = \ell+j \mid D=d ] \notag \\
  =&
{\textstyle 
 \Pr[ \max_i \tilde{\delta}_i - \min_i \tilde{\delta}_i = j \mid D = d ]  }
\notag \\
  =&
{\textstyle 
 \sum_{s=0}^{s_m-j}
 \Pr[ \min_i \tilde{\delta}_i = s, \max_i \tilde{\delta}_i = s+j \mid D = d ] }.
\end{align*}
Since $\tilde{\delta}_1, \dots, \tilde{\delta}_d$ are chosen in independent of each other according to the shift distribution $\Delta (x)$,
Lemma \ref{lem:min_max} gives
\begin{align*}
 &\Pr[\min_i \tilde{\delta}_i = s, \max_i \tilde{\delta}_i = t \mid D = d] \\
  &\quad=
 \Delta_{[s,t]}^d - \Delta_{[s+1,t]}^d - \Delta_{[s,t-1]}^d + \Delta_{[s+1,t-1]}^d.
\end{align*}
Therefore, we get
\begin{align}
 &{\textstyle \sum_{j=0}^{s_m} j \Pr[L = \ell+j \mid D = d] } 
   \notag \\
  =&
 {\textstyle \sum_{j=0}^{s_m} j \sum_{s=0}^{s_m-j}
 \Pr[ \min_i \tilde{\delta}_i = s, \max_i \tilde{\delta}_i = s+j \mid D = d ]}
   \notag \\
  =&
 {\textstyle \sum_{s=0}^{s_m} \sum_{t=s}^{s_m} (t-s)
 \Pr[ \min_i \tilde{\delta}_i = s, \max_i \tilde{\delta}_i = t \mid D = d ] }
   \notag \\
  =&
 {\textstyle \sum_{s=0}^{s_m-1} \sum_{t=s}^{s_m} (t-s) \left( \Delta_{[s,t]}^d - \Delta_{[s+1,t]}^d - \Delta_{[s,t-1]}^d + \Delta_{[s+1,t-1]}^d \right)}
 \notag \\
  =&
 {\textstyle
   s_m
  -\sum_{s=1}^{s_m} \Delta_{[s,s_m]}^d
  -\sum_{t=0}^{s_m-1} \Delta_{[0,t]}^d }.  
 \label{eq:app_a2}
\end{align}
(Note that we use 
$\sum_{t=a}^b (A_t - A_{t-1}) = A_b - A_{a-1}$ and 
$\sum_{t=a}^b t(A_t - A_{t-1})  = b A_b -\sum_{t=a}^{b-1} A_{t} - a A_{a-1} $
at the last equality.)
By combining Eqs.~\eqref{eq:app_a1} and \eqref{eq:app_a2}, we obtain
\begin{align*}
 \mathbb{E}[L]
  &=
 {\textstyle 
 \ell + \sum_{d} \Omega_d 
 \left(
   s_m
  -\sum_{s=1}^{s_m} \Delta_{[s,s_m]}^d
  -\sum_{t=0}^{s_m-1} \Delta_{[0,t]}^d  
   \right)} \\
  &=
 {\textstyle
 \ell  + s_m
 - \sum_{s=1}^{s_m}  \Omega( \Delta_{[s,s_m]} )
 - \sum_{t=0}^{s_m-1}\Omega( \Delta_{[0,t]} ) }.
\end{align*}
This leads Proposition \ref{lem:1}.
\hfill\QED

\bibliographystyle{IEEEtran}
\bibliography{IEEEabrv,nozaki_bib}

\end{document}